\documentclass[a4paper,11pt]{article}
\usepackage[utf8]{inputenc}
\usepackage[T1]{fontenc}
\usepackage[english]{babel}
\usepackage[width=160mm,height=247mm]{geometry}
\usepackage{amsmath, amsfonts, amssymb, bm}
\usepackage[pdftex]{graphicx}
\usepackage{lmodern, indentfirst, cite}
\usepackage[squaren]{SIunits}
\usepackage[textfont={footnotesize},labelfont={color=blue,bf,sf},labelsep=endash]{caption}
\usepackage[colorlinks=true,linkcolor=blue,citecolor=blue,urlcolor=blue]{hyperref}

\usepackage[affil-sl]{authblk}
\title{\textbf{\Large Local quantum uncertainty of a two-qubit XY Heisenberg model with different Dzyaloshinskii-Moriya couplings}}
\author[a]{Younes Moqine
}
\author[b]{Brahim Adnane
}
\author[b]{Abdelhadi Belouad
}
\author[a]{Soufiane Belhouideg
}
\author[b,c]{Rachid Houça\thanks{r.houca@uiz.ac.ma}}
\affil[a]{Research Laboratory of Physics and Engineers Sciences, Team of Applied Physics and New Technologies, Polydisciplinary Faculty, Sultan Moulay Slimane University, Béni Mellal, Morocco}
\affil[b]{LPMC. Laboratory, Theoretical Physics Group, Faculty of Sciences, Choua\"ib Doukkali University, PO Box 20, 24000 El Jadida, Morocco}
\affil[c]{LPTHE. Laboratory, Theoretical Physics and High Energy, Faculty of Sciences, Ibn Zohr University, PO Box 8106, Agadir, Morocco}
\date{}

\def\Tr {{\rm Tr}}
\newcommand{\beq}{\begin{equation}}
\newcommand{\eeq}{\end{equation}}
\newcommand{\lb}{\label}
\DeclareMathOperator{\un}{\emph{I}}
\begin{document}
\begin{titlepage}
	\newgeometry{width=175mm, height=247mm}
    \maketitle
    \thispagestyle{empty}
    \vspace{3cm}
\begin{abstract}
This study investigates the local quantum uncertainty (LQU) of a two--qubit Heisenberg XY chain with different directions of Dzyaloshinskii--Moriya (DM) interactions. The DM interaction parameters and coupling coefficient $J$ are demonstrated to be beneficial in managing correlation. The DM interaction's x--axis parameter has more influence on correlation than the DM interaction's z--axis. As a result, adjusting the direction of the DM interaction may be capable of producing a more efficient operation to improve the correlation.
\end{abstract}

\vspace{2cm}



\end{titlepage}

\section{Introduction}
 Entanglement is an entirely quantum correlation between the parts of a multipartite quantum system that has no classical equivalent. It is acknowledged as a basic physical resource that may be used in various beneficial works in quantum information science \cite{Hor,Niel}. Recently, everyone assumed that quantum correlations were intimately tied to quantum entanglement and that manipulating quantum information was often done in the context of entanglement and separability \cite{Hor}. However, various investigations have demonstrated that entanglement is not the only correlation suitable for implementing quantum protocols and that certain separable states may perform better than their classical counterparts \cite{Fer,Mod}. These efforts have resulted in the invention of a new generation of quantifiers that can identify unclassical correlations beyond entanglement \cite{Mod,Hend,Olli,Dak}. Studying correlations in quantum systems is not confined to linking them to practical applications. Quantum information theory techniques have also been demonstrated to be effective in the investigation of condensed matter systems \cite{Oste}.

Condensed matter systems’ quantum entanglement is a significant field, as is well known. Various studies on quantum entanglement have been accomplished on the thermal equilibrium states of spin chains at a fixed temperature \cite{Amesen,WangX.G,ZhouL,ZhangG.F,Guo1}. Moreover, two-qubit quantum correlations with the DM interaction receive much attention from researchers \cite{Houca1,Tufarelli,YaoY1,Song23,Yao2,Liu23}. In addition, the Heisenberg model was used to study entanglement. A number of important works were produced, including the isotropic Heisenberg XX model \cite{Wang X G33,Wang X G34}, the XXX model \cite{Houca2,Nielsen,Arnesen}, the anisotropic Heisenberg XY model \cite{Kamta G L,Wang X G35}, the completely anisotropic Heisenberg XYZ model \cite{Zhou L36}.

Local quantum uncertainty captures exclusively quantum correlations and excludes classical correlations. This measure is of the quantum discord type, but it has the benefit of not requiring a complex optimization technique over measurements. This measure was first developed for bipartite quantum systems, and there is no closed formula for $2\otimes D$ systems. A discord-like measurement named local quantum uncertainty \cite{Gir} was recently introduced. This measurement uses skew information from a single local measurement \cite{Luo,Wig,Luoo}. For $2\otimes D$ bipartite quantum systems, the LQU measure has a closed formula \cite{mez}. Later, other researchers attempted to study local quantum uncertainty for the orthogonally invariant class of states \cite{Sen}. This measurement was also examined in the context of quantum phase transitions \cite{Kar,Coul}. It also explored the link between local quantum uncertainty and quantum Fisher information in a non-Markovian environment \cite{Wu}. Some authors have recently examined LQU under different decoherence models and produced early findings for three qubits \cite{Sla1,Sla2,Sla3,Sla4}.

Our motivation in this work is to examine the impact of the various DM interaction components on LQU. To this end, we will explore a two-qubit Heisenberg XY chain with an x--axis and z--axis DM interaction in order to investigate LQU's thermal entanglement. Our centrosymmetric Hamiltonian matrix of x--axis  DM interaction will be transformed into the X--form via a double Hadamard transformation to examine the effects of the two DM interactions on the entanglement.

The structure of our article is as follows. We introduce the basics of the LQU in section {\color{blue}2} and offer an analytical expression of an X–types matrix. In Section {\color{blue}3}, we present the two--qubit anisotropic Heisenberg XY chain Hamiltonian with the z--axis DM interaction, evaluate the LQU of this system and examine the effects of parameters on the entanglement in the ground state and thermal state. Section {\color{blue}4} studies the two-qubit Heisenberg XY chain model using the x--axis DM interaction. Then, in Section {\color{blue}5}, we compared the effects of the two DM directions on the entanglement. Finally, the discussion in Section {\color{blue}6} completes the article.
\section{LQU  of a two--qubit X states}
It is reasonable to think of the idea of LQU as a way to quantify non-classical correlations in multipartite systems. This difference was put up in an effort to measure the lowest quantum uncertainty that may be introduced into a quantum state by measuring one local observable \cite{Girolami}. Considering the density matrix
\begin{equation} \label{varo}
\varrho=\left(
\begin{array}{cccc}
\rho_{11} & 0 & 0 &  \rho_{14} \\
0 & \rho_{22} & \rho_{23} & 0 \\
0 & \overline{\rho_{23}} & \rho_{33} & 0 \\
 \overline{\rho_{14}} & 0 & 0 & \rho_{44}
\end{array}
\right)
\end{equation}
For two subsystems $A$ and $B$,  The LQU in relation to the $A$ is defined by
\begin{equation}\label{LQU}
\mathcal{U}(\rho_{AB}) := \min_{\varpi_{A}^{\mu} \otimes
\un_B} \mathcal{I}\Big(\rho_{AB}, \varpi_{A}^{\mu}\otimes
\un_B\Big).
\end{equation}
where $\varpi_{A}^{\mu}\otimes \un_B$ denotes a local observable, with $\varpi_{A}^{\mu}$ is a Hermitian operator on subsystem $A$ which is an adequate uncertainty quantifier with spectrum $\mu$ and $\un_B$ is the identity matrix operating on $B$. The minimal is optimized overall local observables on $A$ with
\begin{equation}\lb{skew}
\mathcal{I}(\chi,  \varpi_{A}^{\mu} \otimes
\un_B):=-\frac{1}{2}{\text{
Tr}}\left(\Big[\sqrt{\chi}, \varpi_{A}^{\mu} \otimes
\un_B\Big]^{2}\right).
\end{equation}
is the Wigner-Yanase skew information \cite{Wig,Luo}. It measures the uncertainty of the observable $\varpi$ for the state $\chi$. We can simply show for pure states $(\chi^2=\chi)$ which the skew information reduces the usual variance formula
\beq\lb{var}
\text{Var}\left(\chi, \varpi\right)=\text{Tr}\left(\chi \varpi^2\right)-\Big(\text{Tr}\big(\chi \varpi\big)\Big)^2.
\eeq
Through the ensemble of all observables operating on the element $A$, a minimization approach is used to accomplish the analytical calculation of the local quantum uncertainty. A closed form for qubit--qudit systems was obtained after optimization in \cite{Wang}. The local quantum uncertainty concerning subsystem $A$ is specifically provided for a qubit by \cite{Girolami}
\begin{equation}\lb{lqu}
 \mathcal{U}(\rho_{AB}) = 1 - \lambda_{\max}\left\{\mathcal{W}_{AB}\right\},
\end{equation}
where $\lambda_{\max}$ is the greatest eigenvalue of the $3\times 3$ symmetric matrix $\mathcal{W}_{AB}$ with
\begin{equation}\label{matrixomega}
 \Big(\mathcal{W}_{AB}\Big)_{lk} \equiv  {\text{
Tr}}\left\{\sqrt{\rho_{AB}}\Big(\sigma_{A}^{l}\otimes
{\un}_{B}\Big)\sqrt{\rho_{AB}}\Big(\sigma_{A}^{k}\otimes {\un}_{B}\Big)\right\}
\end{equation}
where $\sigma_{A}^{l,k} (l, k = x, y, z)$ indicates the Pauli matrices of the subsystem $A$. The LQU has been shown to meet all of the physical proprieties for a quantum correlations measurement \cite{Girolami}. The LQU has several fascinating features. Among them are its invariance under any local unitary operations. Furthermore, LQU is a dependable discord-like measure with geometrical importance in terms of Hellinger distance \cite{Luo,Girolami}.\\
The phase factors $e^{i\alpha_{14}}$ and $e^{i\alpha_{23}}$ may be omitted from the off--diagonal components since the LQU is invariant under local unitary transformations. Using the following local unitary transformations \cite{Kedif,Kedif1}
\begin{equation*}
|0\rangle_A=\exp\Big(-\frac{i}{2}\big(\alpha_{14}+\alpha_{23}\big)\Big)|0\rangle_A \ \ \ {\text{and}} \ \ \ |0\rangle_B=\exp\Big(-\frac{i}{2}\big(\alpha_{14}-\alpha_{23}\big)\Big)|0\rangle_B,
\end{equation*}
the density matrix $\varrho$ becomes
 \begin{equation}\label{varrhop}
\varrho\longrightarrow\varrho^{\prime} =  \left(
\begin{array}{cccc}
\rho_{11} & 0 & 0 & \vert \rho_{14} \vert \\
0 & \rho_{22} & \vert \rho_{23} \vert & 0 \\
0 & \vert \rho_{23} \vert & \rho_{33} & 0 \\
\vert \rho_{14} \vert & 0 & 0 & \rho_{44}
\end{array}
\right).
\end{equation}
The Fano--Bloch decomposition of the state $\rho$ reads
\beq
\rho={1\over4}\sum_{\alpha,\beta}R_{\alpha,\beta}\sigma_{\alpha}\otimes\sigma_{\beta}
\eeq
where the correlation matrix $R_{\alpha,\beta}$ are given by $R_{\alpha,\beta}=\Tr\left(\rho\sigma_{\alpha}\otimes\sigma_{\beta}\right)$, with $\alpha, \beta=0, 1, 2, 3$. Explicitly, they write
\begin{eqnarray}\lb{fano}
R_{00} &=& \Tr(\rho)=1 \\  \nonumber
R_{11} &=& 2 \left(|\rho _{23}| +|\rho _{41}| \right) \\ \nonumber
R_{22} &=& 2 \left( |\rho _{23}| - |\rho _{41}| \right) \\ \nonumber
R_{33} &=& 1-2 \left(\rho _{22}+\rho _{33}\right) \\ \nonumber
R_{03} &=&1 -2 \left(\rho _{22}+\rho _{44}\right) \\ \nonumber
R_{30} &=& 1-2 \left(\rho _{33}+\rho _{44}\right)  \nonumber
\end{eqnarray}
To determine the LQU defined by \eqref{lqu}, one
should compute the eigenvalues of the matrix $\mathcal{W}_{AB}$ \eqref{matrixomega}. Explicitly, they are given by \cite{Jebli}
\begin{eqnarray}\lb{10}
 \omega_1 &=&\sqrt{\left(2 \sqrt{d_1}+t_1\right) \left(2 \sqrt{d_2}+t_2\right)}+\frac{R_{03}^2+R_{11}^2-R_{22}^2-R_{30}^2}{4 \sqrt{\left(2 \sqrt{d_1}+t_1\right) \left(2 \sqrt{d_2}+t_2\right)}} \\ \nonumber
 \omega_2 &=&\sqrt{\left(2 \sqrt{d_1}+t_1\right) \left(2 \sqrt{d_2}+t_2\right)}+\frac{R_{03}^2-R_{11}^2+R_{22}^2-R_{30}^2}{4 \sqrt{\left(2 \sqrt{d_1}+t_1\right) \left(2 \sqrt{d_2}+t_2\right)}} \\ \nonumber
 \omega_3 &=& \frac{1}{2} \left(2 \left(\sqrt{d_1}+\sqrt{d_2}\right)+1\right)+\frac{1}{8} \left(\frac{\left(R_{03}+R_{03}\right){}^2-\left(R_{22}-R_{11}\right){}^2}{2 \sqrt{d_1}+t_1}+\frac{\left(R_{03}-R_{30}\right){}^2-\left(R_{11}+R_{22}\right){}^2}{2 \sqrt{d_2}+t_2}\right)
\end{eqnarray}
where
\begin{eqnarray}\lb{11}
  t_1 &=& \rho _{11}+\rho _{44} \\ \nonumber
  t_2 &=& \rho _{22}+\rho _{33} \\ \nonumber
  d_1 &=& \rho _{11} \rho _{44}-\rho _{14} \rho _{41} \\ \nonumber
  d_2 &=& \rho _{22} \rho _{33}-\rho _{23} \rho _{32}  \nonumber
\end{eqnarray}
It is must be noticed that $R_{11}\geq R_{22}$. This implies that $\omega_1>\omega_2$. As a result of the equation \eqref{lqu}, the LQU measuring the pairwise quantum correlation in the state $\varrho$ reads as
\begin{equation}\lb{lqu1}
 \mathcal{U}(\varrho) = 1 - {\max}\left\{\omega_1, \omega_3\right\}.
\end{equation}
Lastly, comparing the eigenvalues $\omega_1$ and $\omega_3$ is an essential aspect of the LQU calculation technique, and the largest one must be found.
\section{XY Heisenberg model with z--axis DM interaction} \lb{sec3}
\subsection{Theoretical model and density matrix}
The Hamiltonian $\mathcal{H}$ for a two-qubit anisotropic Heisenberg XY model with  z--axis DM interaction is written by
\beq\lb{fr}
\mathcal{H}= J\left(\sigma_x^1\sigma_x^{2}+\Delta\sigma_y^{1}\sigma_y^{2}\right)+D_z\left(\sigma_x^{1}\sigma_y^{2}-\sigma_y^{1}\sigma_x^{2}\right)
\eeq
where $D_z$ is the z--axis DM interaction, $\sigma_i(i = x, y, z)$ denotes the spin--1/2 Pauli matrices, the dimensionless factor $\Delta\in [0, 1]$ is the anisotropy parameter along y--direction. It should be noted that $J$ represents the coupling between the spin chains. For $J > 0$, the chain is antiferromagnetic, and for   $J< 0$, the chain is ferromagnetic.
As seen below, the Hamiltonian \eqref{fr} may be represented under its matrix form in the usual computational basis $|00>$, $|01 >$, $|10 >$, $|11 >$ by
\beq\lb{1}
\mathcal{H}=\left(
\begin{array}{cccc}
 0 & 0 & 0 & J(1-\Delta ) \\
 0 & 0 & J(\Delta +1)+2 i D_z & 0 \\
 0 & J(\Delta +1)-2 i D_z & 0 & 0 \\
 J(1-\Delta ) & 0 & 0 & 0 \\
\end{array}
\right)
\eeq
The solution of the eigenvalue equation leads to the eigenvalues, which are listed below
\begin{eqnarray}\lb{22}
\epsilon_{1,2} &=&\pm J(\Delta -1)\\ \nonumber
\epsilon_{3,4} &=&  \pm\Omega
\end{eqnarray}
and the eigenvectors of the system
\begin{eqnarray}\lb{psi}
  |\varphi_1\rangle &=& -\frac{1}{\sqrt{2}}|00\rangle+\frac{1}{\sqrt{2}}|11\rangle \\ \nonumber
  |\varphi_2\rangle &=& \frac{1}{\sqrt{2}}|00\rangle+\frac{1}{\sqrt{2}}|11\rangle \\ \nonumber
  |\varphi_3\rangle &=& \frac{e^{i \theta }}{\sqrt{2}}|01\rangle+\frac{1}{\sqrt{2}}|10\rangle \\ \nonumber
  |\varphi_4\rangle &=& -\frac{e^{i \theta }}{\sqrt{2}}|01\rangle+\frac{1}{\sqrt{2}}|10\rangle
\end{eqnarray}
where the quantities $\Omega$ and $\theta$  are defined by
\begin{eqnarray}
 \Omega &=& \sqrt{4 D_z^2+(\Delta +1)^2 J^2}\\ \nonumber
 \theta &=& \arctan\left(\frac{2 D_z}{J(\Delta +1)}\right)
\end{eqnarray}
Prior to investigating the thermal quantum entanglements at finite temperatures, we must first address the entanglement of the system's ground state at absolute zero temperature. Given that the energies rely on the exchange coupling $J$, it is conceivable to think that ground state entanglement arises in both antiferromagnetic and ferromagnetic cases.\\
In the case $J > 0$, using the equation (\ref{22}), we can write the ground state energies as
\begin{eqnarray}
\epsilon_{1}&= J(\Delta -1),&\qquad \text{if} \quad  J>\frac{\Omega}{(1-\Delta)},\\ \nonumber
\epsilon_{4}&=-\Omega,&\qquad \text{if} \quad J<\frac{\Omega}{(1-\Delta)}.
\end{eqnarray}
Consequently, when the condition $J>\frac{\Omega}{(1-\Delta)}$ is fulfilled, $|\varphi_1\rangle$ becomes the ground state which is an entangled state but, when $J<\frac{\Omega}{(1-\Delta)}$ is verified, the ground state turns into the ground state $|\varphi_4\rangle$. In addition, $|\varphi_1\rangle$ and $|\varphi_4\rangle$ are maximally entangled states, thus conducting to the maximum of $LQU=1$ is achieved at low temperatures.\\
In the case J < 0, the ground state energies are written by
\begin{eqnarray}
\epsilon_{2}&= -J(\Delta -1),&\qquad \text{if} \quad  J<-\frac{\Omega}{(1-\Delta)},\\ \nonumber
\epsilon_{4}&=-\Omega,&\qquad \text{if} \quad J>-\frac{\Omega}{(1-\Delta)}.
\end{eqnarray}
where the maximally entangled states $|\varphi_2\rangle$ and $|\varphi_4\rangle$ that corresponds to the maximum local quantum uncertainty $LQU=1$.\\
After obtaining the spectrum of our system, it is easy to calculate the thermal density, which is essential for performing measurements of the examined system's entanglement at temperature $T$. Indeed the expression of the $\rho(T)$ is given by
\beq
\rho(T)={1\over\mathbb{Z}}e^{-\beta \mathcal{H}}
\eeq
such that the canonical partition function $\mathbb{Z}$ is written by
\beq
\mathbb{Z}=\Tr e^{-\beta \mathcal{H}}
\eeq
where $k_B$ is the constant of the Boltzmann, for convenience, it is considered as unity in the next. The temperature's inverse is represented by the parameter $\beta=1/T$. The spectrum of the Hamiltonian \eqref{1} allows expressing the thermal density $\rho(T)$ as
\beq\lb{3}
\rho(T)={1\over\mathbb{Z}}\sum_{l=1}^{4}e^{-\beta \epsilon_l}|\phi_l\rangle\langle\phi_l|
\eeq
By inserting \eqref{22} and \eqref{psi} in the equation \eqref{3}, we may derive the system's density matrix in the standard computational basis
\beq\lb{4}
\rho(T)={1\over\mathbb{Z}}\left(
\begin{array}{cccc}
 \cosh (\beta  J(\Delta -1)) & 0 & 0 & \sinh (\beta  J(\Delta -1)) \\
 0 & \cosh (\beta  \Omega ) & -e^{i \theta } \sinh (\beta  \Omega ) & 0 \\
 0 & -e^{-i \theta } \sinh (\beta  \Omega ) & \cosh (\beta  \Omega ) & 0 \\
 \sinh (\beta  J(\Delta -1)) & 0 & 0 & \cosh (\beta  J(\Delta -1)) \\
\end{array}
\right)
\eeq
whereby the partition function is defined explicitly by
\beq
\mathbb{Z}=2 \cosh (\beta  \Omega )+2 \cosh (\beta  J(\Delta -1))
\eeq
Because $\rho(T)$ represents a thermal state, the quantum correlations it generates are referred to as thermal quantum correlations. The following section will quantify the quantum correlations in the aforementioned two-qubit system as a function of the system's characteristics, which include the x--axis parameter $D_x$ and the z--axis $D_z$  DM interaction parameters, as well as the temperature $T$ and coupling coefficient $J$.
\subsection{Local quantum uncertainty}
The density matrix \eqref{4} clearly has the X--type form. Consequently, the LQU may be easily calculated using the findings mentioned earlier. To get the LQU expression, we must determine the eigenvalues of the matrix $\mathcal{W}_{AB}$. The non-vanishing matrix correlation components $R_{\mu,\nu}$ that emerge during the Fano-Bloch decomposition of the density matrix $\rho(T)$ must first be evaluated. The outcome of combining the equations \eqref{fano}, \eqref{11}, and the density matrix components \eqref{4} are
\begin{eqnarray}\lb{R}
R_{00} &=& \Tr(\rho)=1 \\ \nonumber
R_{11} &=& \frac{\left| \sinh ( \beta(\Delta -1)J)\right| +\sinh (\beta  \Omega )}{\cosh (\beta  \Omega )+\cosh (\beta(\Delta -1)J)} \\ \nonumber
R_{22} &=& \frac{\sinh (\beta  \Omega )-\left| \sinh ( \beta(\Delta -1)J)\right| }{\cosh (\beta  \Omega )+\cosh (\beta(\Delta -1)J)} \\ \nonumber
R_{33} &=& 1-\frac{2 \cosh (\beta  \Omega )}{\cosh (\beta  \Omega )+\cosh (\beta(\Delta -1)J)} \\ \nonumber
R_{03} &=&R_{03}=0 \nonumber
\end{eqnarray}
and the  quantities $t_1$ and $d_1$ are given by
\begin{eqnarray}\lb{t1}
  t_1 &=& 1-t_2=\frac{1}{\cosh (\beta  \Omega ) \text{sech}(\beta(\Delta -1)J)+1} \\ \nonumber
  d_1 &=&d_2= \frac{1}{4 (\cosh (\beta  \Omega )+\cosh (\beta(\Delta -1)J))^2} \\ \nonumber
\end{eqnarray}
Now we have all the ingredients to obtain the eigenvalues of the matrix $\mathcal{W}_{AB}$, just replace the different quantities in \eqref{10} and \eqref{11} by their expressions found in \eqref{R} and  \eqref{t1} and therefore, the LQU may be expressed in terms of $ \omega _1$ and $ \omega _3$ as follows:
\begin{equation}\lb{lqu1}
 \mathcal{LQU}= 1 - {\max}\left\{\omega_1, \omega_3\right\}.
\end{equation}
We examine the LQU of Eq. \eqref{lqu1} in order to identify how the entanglement has changed. After deriving the LQU equation, which depends on the temperature $T$, the z--axis DM interaction $D_z$, the y--direction anisotropy parameter $\Delta$, and the coefficient $J$. Therefore, we have all the elements needed to examine how the recommended system behaves concerning the quantities specified before. We may learn about the functions of other factors and the variations in entanglement by setting the other parameters.

\begin{figure}[!h]
  \centering
  \includegraphics[width=6.5in]{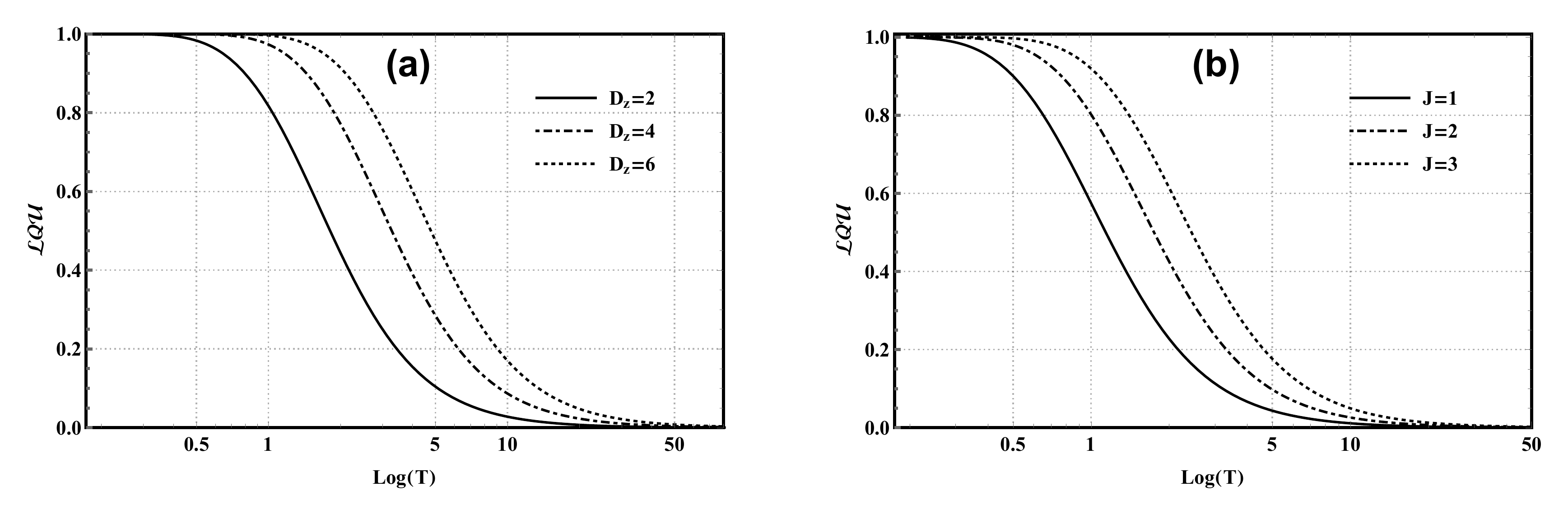}
  \caption{: The LQU versus logarithmic temperature. (a): for various values of z-component parameter $D_z$ with $J=1$. (b): for various values of $J$ with $D_z=1$ and $\Delta$ is set equal $0.5$. All the parameters are dimensionless}\label{f1}
\end{figure}
In Fig. (\ref{f1}.a), the LQU is plotted versus logarithmic temperature $Log(T)$ for different z--axis parameter $D_z$ where the coupling constants $J = 1$  and $\Delta=0.5$. The figure shows three significant remarks. In the First one, we observe that the thermal LQU decreases when the temperature increases. In the second one, the LQU remain equal to one at a low--temperature regime even $T$ raise. In this case, the system's ground state is $|\varphi_4\rangle$ which is the maximally entangled state. In addition, we note that the LQU decreases rapidly for the value of $Log(T)=0.4$ (see the solid black curve). Long--lasting entanglement in low temperatures requires large values of z--component of DM interaction. The last one, in the high--temperature regime, one can notice critical temperatures $T_c$ beyond which the LQU disappears. The threshold temperature $T_c$ is influenced by the DM interaction via the parameter $D_z$, such that $T_c$ and $D_z$ rise proportionally. From Fig. (\ref{f1}.b), we notice that the impact of $J$ is similar to the influence of $D_z$ on the LQU. For this, the remarks of a figure (\ref{f1}.a) remain the same.

\begin{figure}[!h]
  \centering
  \includegraphics[width=3.3in]{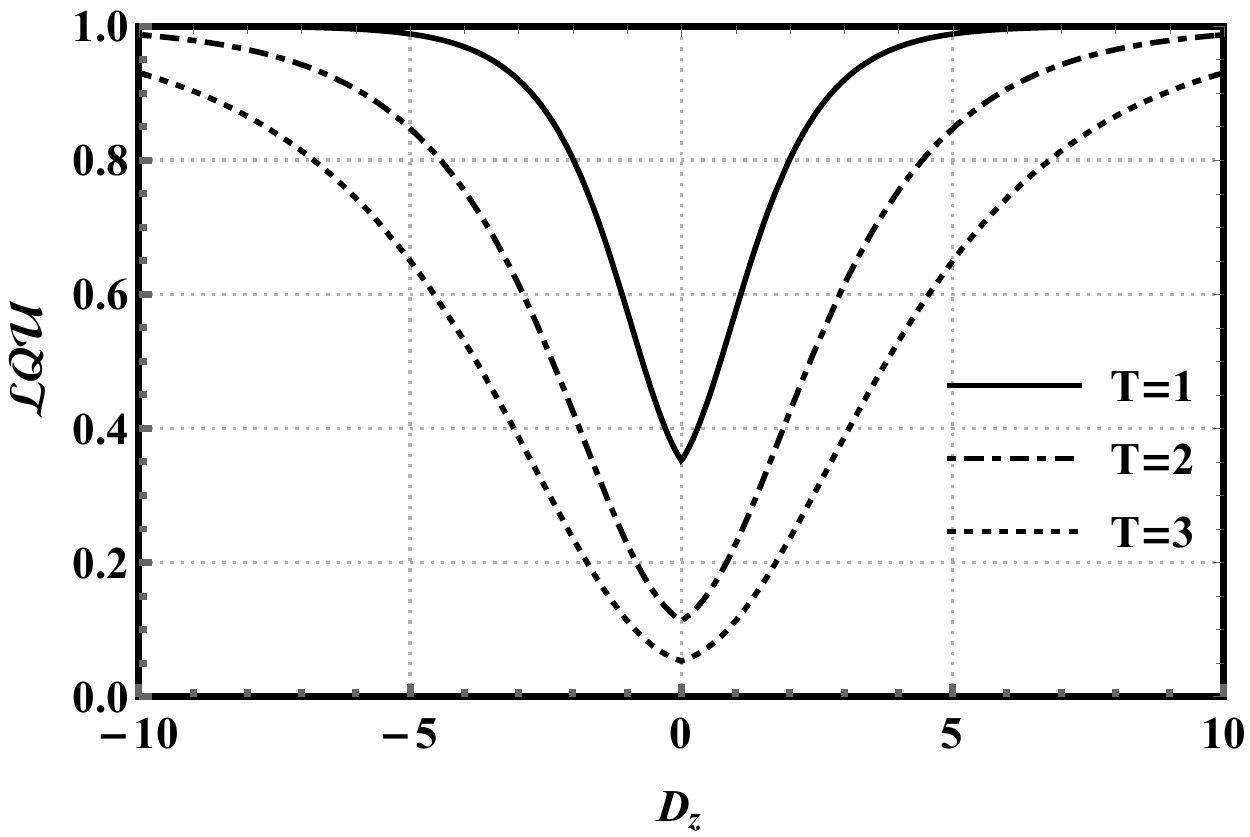}
  \caption{: The LQU versus $D_z$ for various values of $T$, with $J=1$ and $\Delta=0.5$.}\label{f3}
\end{figure}
In Fig. (\ref{f3}), we graph the LQU in terms of the z--axis $D_z$ for different temperature values such as $T= 1, 2, 3$. The first finding is that the LQU is symmetrical for $D_z=0$ and goes to a fixed value $LQU = 1$ for the high absolute value of the z--axis of DM interaction even as temperature increases. Furthermore, for $D_z = 0$, we see that the LQU has a minimum for any temperature value, which diminishes as the temperature decreases and vice versa, as seen in Fig. (\ref{f1}). Then we may deduce that the DM interaction impacts the correlation, i.e., the system becomes increasingly entangled due to the high values of $|D_z|$.

\section{ XY HEISENBERG MODEL WITH x--axis DM INTERACTION} \lb{sec4}
\subsection{Theoretical model and density matrix}
In this part, we take into consideration the same Hamiltonian as before, but we reverse the direction of DM interaction such that
\beq\lb{frr}
\mathcal{H'}= J\left(\sigma_x^1\sigma_x^{2}+\Delta\sigma_y^{1}\sigma_y^{2}\right)+D_x\left(\sigma_y^{1}\sigma_z^{2}-\sigma_z^{1}\sigma_y^{2}\right)
\eeq
where $D_x$ is the x--axis DM interaction, J, $\Delta$ and $\sigma_i(i = x, y, z)$ are the same as those in section (\ref{sec3}). In the usual computational basis $|00>$, $|01 >$, $|10 >$, $|11 >$  the Hamiltonian \eqref{frr} can be rewritten as
\beq
\mathcal{H'}=\left(
\begin{array}{cccc}
 0 & i D_x & -i D_x & J(1-\Delta ) \\
 -i D_x & 0 & J(\Delta +1) & i D_x \\
 i D_x & J(\Delta +1) & 0 & -i D_x \\
 J(1-\Delta ) & -i D_x & i D_x & 0 \\
\end{array}
\right)
\eeq
solving the eigenvalue equation of $\mathcal{H'}$ leads to the eigenstates
\begin{eqnarray}\lb{psi1}
  |\phi_1\rangle &=& \frac{1}{\sqrt{2}}\left(|01\rangle+|10\rangle\right) \\ \nonumber
  |\phi_2\rangle &=& \frac{1}{\sqrt{2}}|00\rangle+\frac{1}{\sqrt{2}}|11\rangle \\ \nonumber
  |\phi_3\rangle &=& \frac{1}{\sqrt{2}}\left(-\frac{1}{\sqrt{\frac{\zeta_1^2}{4}+1}}|00\rangle+\frac{i \zeta_1}{\sqrt{\zeta_1^2+4}}|01\rangle-\frac{i \zeta_1}{\sqrt{\zeta_1^2+4}}|10\rangle+\frac{1}{\sqrt{\frac{\zeta_1^2}{4}+1}}|11\rangle \right)\\ \nonumber
  |\phi_4\rangle &=&\frac{1}{\sqrt{2}}\left(-\frac{1}{\sqrt{\frac{\zeta_2^2}{4}+1}}|00\rangle-\frac{i \zeta_2}{\sqrt{\zeta_2^2+4}}|01\rangle+\frac{i \zeta_2}{\sqrt{\zeta_2^2+4}}|10\rangle+\frac{1}{\sqrt{\frac{\zeta_2^2}{4}+1}}|11\rangle \right)
\end{eqnarray}
with corresponding eigenvalues
\begin{eqnarray}\lb{cc}
\epsilon_{1,2}' &=& J( 1\pm\Delta)\\ \nonumber
\epsilon_{3,4}' &=&  -J\pm\Omega_1
\end{eqnarray}
where $\Omega_1 =\sqrt{4 D_x^2+\Delta ^2 J^2}$, and $\zeta_{1,2}=\frac{\Omega_1 \mp\Delta  J}{D_x}$. The equation \eqref{cc} shows that the  energies depends on the sign of $J$, then for $J<0$, the ground state energies are given by
\begin{eqnarray}\lb{WWW}
\epsilon_{2}' &= J(1-\Delta),&\qquad \text{if} \quad  J<-\frac{\Omega_1}{(2-\Delta)},\\ \nonumber
\epsilon_{4}' &= -J-\Omega_1,&\qquad \text{if} \quad J>-\frac{\Omega_1}{(2-\Delta)}.
\end{eqnarray}
While if $J>0$, it is given by
\begin{eqnarray}\lb{XXXX}
\epsilon_{4}' &=& -J-\Omega_1.
\end{eqnarray}
As a result, $|\phi_4\rangle$ is always the ground state pertinent to the antiferromagnetic system. It is the ground state for ferromagnetic systems as well, but only when the condition $J>-\frac{\Omega_1}{(2-\Delta)}$ is fulfilled.\\
Inserting the Eqs. \eqref{psi1} and \eqref{cc} into Eq.\eqref{3} the density matrix can be written by
\beq
\rho'(T)=\left(
\begin{array}{cccc}
 a & \mu & \nu & b \\
 \nu & c & d & \mu \\
 \mu & d & c & \nu \\
 b & \nu & \mu & a \\
\end{array}
\right)
\eeq
such as their elements are represented by the expressions
\begin{eqnarray}\lb{23}
a &=&\frac{e^{\beta  J} \left(\cosh (\beta  \Omega_1 )-\frac{\Delta  J \sinh (\beta  \Omega_1 )}{\Omega_1 }\right)+e^{\beta  J(\Delta -1)}}{2 \mathcal{Z'}} \\ \nonumber
b &=& \frac{e^{\beta J (\Delta -1)}-e^{\beta  J} \left(\cosh (\beta  \Omega_1 )-\frac{\Delta  J \sinh (\beta  \Omega_1 )}{\Omega_1 }\right)}{2 \mathcal{Z'}} \\ \nonumber
c &=& \frac{e^{\beta  J} \left(\cosh (\beta  \Omega_1 )+\frac{\Delta  J \sinh (\beta  \Omega_1 )}{\Omega_1 }\right)+e^{-\beta J  (\Delta +1)}}{2 \mathcal{Z'}} \\ \nonumber
d &=&\frac{e^{-\beta J (\Delta +1)}-e^{\beta  J} \left(\cosh (\beta  \Omega_1 )+\frac{\Delta  J \sinh (\beta  \Omega_1 )}{\Omega_1 }\right)}{2 \mathcal{Z'}}\\ \nonumber
\mu &=&-\frac{i D_x e^{\beta  J} \sinh (\beta  \Omega_1 )}{\Omega_1  \mathcal{Z'}} \\ \nonumber
\nu &=&\frac{i D_x e^{\beta  J} \sinh (\beta  \Omega_1 )}{\Omega_1  \mathcal{Z'}}
\end{eqnarray}
where the partition function $\mathcal{Z'}=2 \left(e^{-\beta J} \cosh (\beta  \Delta  J)+e^{\beta  J} \cosh (\beta  \Omega_1 )\right)$. At this stage, the determination of LQU is complicated. For this reason, we will make a unitary transformation of the density matrix to render the density matrix in the X--form. To do this, we will apply a double Hadamard transformation to our centrosymmetric density matrix by the unitary matrix
\beq
H=\frac{1}{\sqrt{2}}\left(
\begin{array}{cc}
 1 & 1 \\
 1 & -1 \\
\end{array}
\right)
\eeq
After a straightforward calculation we obtain
\beq\lb{44}
\rho'_H=\frac{e^{-\beta J}}{\mathcal{Z'}}\left(
\begin{array}{cccc}
  \cosh (\beta  \Delta  J) & 0 & 0 &  \sinh (\beta  \Delta  J) \\
 0 & e^{2\beta  J} \cosh (\beta  \Omega_1 ) & -\frac{e^{2\beta  J} \sinh (\beta  \Omega_1 ) \left(\Delta J+2 i D_x\right)}{\Omega_1 } & 0 \\
 0 & -\frac{e^{2\beta  J} \sinh (\beta  \Omega_1 ) \left(\Delta  J-2 i D_x\right)}{\Omega_1 } & e^{2\beta  J} \cosh (\beta  \Omega_1 ) & 0 \\
  \sinh (\beta  \Delta  J) & 0 & 0 &  \cosh (\beta  \Delta  J) \\
\end{array}
\right)
\eeq
The density matrix \eqref{44} clearly have now the X--type form, which is very easy to calculate the LQU for the studied system.
\subsection{Local quantum uncertainty}
To get the expression of LQU, we will determine the eigenvalues of the $\mathcal{W}_{AB}$ matrix. One must first evaluate the non--vanishing matrix correlation components $R'_{\mu,\nu}$ that emerge during the Fano--Bloch decomposition of the density matrix $\rho_H$. When the equations \eqref{fano}, \eqref{11} and the density matrix elements \eqref{4} are combined, the result is
\begin{eqnarray}\lb{R}
R'_{00} &=& \Tr(\rho)=1 \\ \nonumber
R'_{11} &=& \frac{\left| \sinh (J \beta  \Delta )\right| +e^{2 \beta  J} \sinh (\beta  \Omega_1 )}{\cosh (\beta  \Delta  J)+e^{2 \beta  J} \cosh (\beta  \Omega_1 )} \\ \nonumber
R'_{22} &=& \frac{e^{2 \beta  J} \sinh (\beta  \Omega_1 )-\left| \sinh (J \beta  \Delta )\right| }{\cosh (\beta  \Delta  J)+e^{2 \beta  J} \cosh (\beta  \Omega_1 )} \\ \nonumber
R'_{33} &=& \frac{2}{e^{2 \beta  J} \cosh (\beta  \Omega_1 ) \text{sech}(\beta  \Delta  J)+1}-1 \\ \nonumber
R'_{03} &=&R'_{30}=0 \nonumber
\end{eqnarray}
and the  quantities $t_1$ and $d_1$ are given by
\begin{eqnarray}\lb{t1}
  t'_1 &=& 1-t'_2=\frac{1}{e^{2 \beta  J} \cosh (\beta  \Omega_1 ) \text{sech}(\beta  \Delta  J)+1} \\ \nonumber
  d'_1 &=& \frac{1}{4 \left(\cosh (\beta  \Delta  J)+e^{2 \beta  J} \cosh (\beta  \Omega_1 )\right)^2} \\ \nonumber
  d'_2 &=& \frac{e^{4 \beta  J}}{4 \left(\cosh (\beta  \Delta  J)+e^{2 \beta  J} \cosh (\beta  \Omega_1 )\right)^2}
\end{eqnarray}
then the LQU for this situation is
\begin{equation}\lb{lqu11}
 \mathcal{LQU'}= 1 - {\max}\left\{\omega'_1, \omega'_3\right\}.
\end{equation}

\begin{figure}[!h]
  \centering
  \includegraphics[width=6.5in]{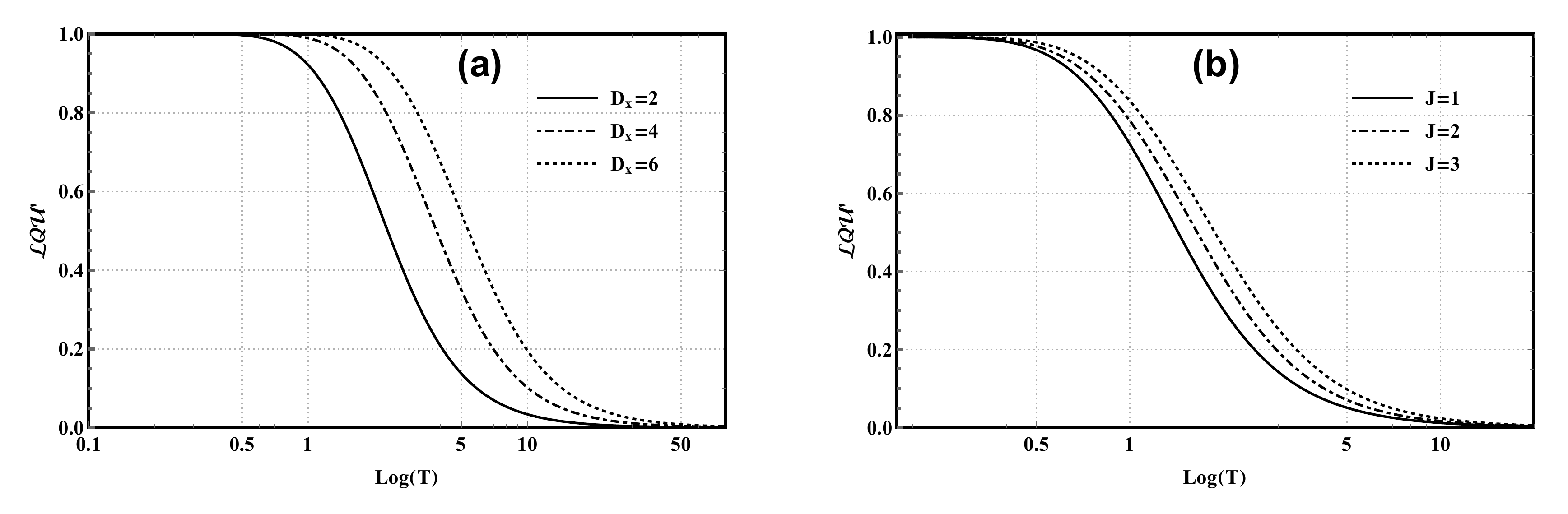}
  \caption{: The LQU' versus $Log(T)$. (a): for various values of x-component parameter $D_x$ with $J=1$. (b): for various values of $J$ with $D_x=1$ and $\Delta$ is set equal $0.5$. All the parameters are dimensionless.}\label{f4}
\end{figure}
Fig.(\ref{f4}.a) shows the LQU' versus Logarithmic temperature for different DM coupling parameter $D_x$ when $J=1$ and $\Delta=0.5$. We observe that at low temperatures and precisely in the interval $0 < Log(T) < 0.5$, the value of LQU' equals one whatever the value of $D_x$. In this case, the ground state of the system is $ |\varphi_4\rangle$, which is the maximally entangled state, then it starts decreasing for small values of $D_x$, and Obviously, increasing temperature will decrease entanglement and increasing $D_x$ will increase entanglement and critical temperature $T_c$. Figure (\ref{f4}.b) demonstrates that the LQU' with respect to $Log(T)$ for different $J$ with the x--component parameter $D_x = 1$ and $\Delta = 0.5$, it is easy to find that increasing $J$ can increase the critical temperature and improve entanglement for a certain temperature. Thus, $D_x$ and $J$ are also effective control parameters for entanglement.

\begin{figure}[!h]
  \centering
  \includegraphics[width=3.5in]{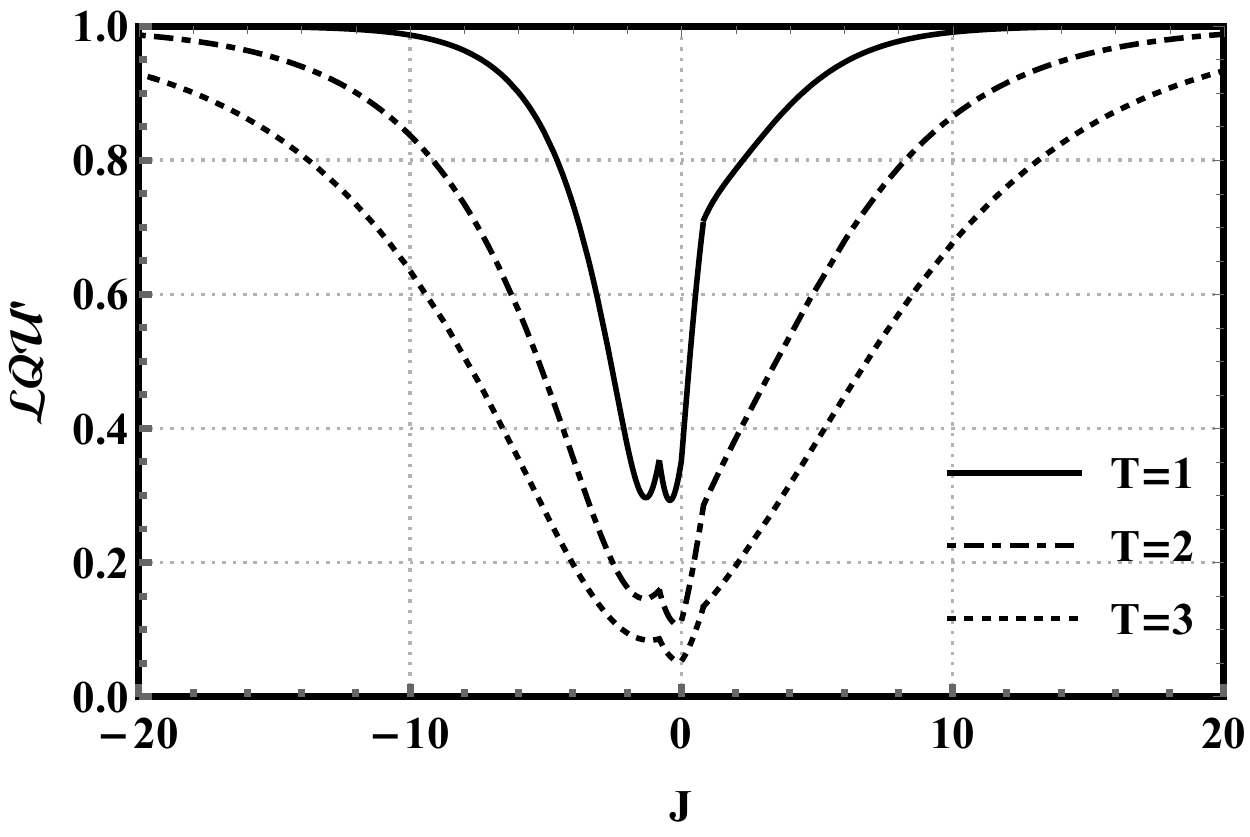}
  \caption{: The LQU' versus $J$ for various values of $T$ with $D_x=1$ and $\Delta=0.5$.}\label{f6}
\end{figure}
Fig.(\ref{f6}) shows the changes of LQU' in terms of interaction parameter $J$ for various values of $T$, where the x-component parameter $D_x = 1$ and $\Delta = 0.5$. As the value of this parameter increases, the amount of LQU' increases and reaches to constant value one both for the systems with ferromagnetic and anti-ferromagnetic nature. It also can be seen the LQU' is decreased by increasing temperature, which is confirmed by fig.(\ref{f4}.b).
\section{Impact of the DM interaction directions on LQU}
We know from sections \ref{sec3} and \ref{sec4} that the x-axis parameter $D_x$ and the z--axis parameter $D_z$ of the DM interaction have comparable properties. They are both efficient correlation control factors; raising them may improve correlation or raise the critical temperature, slowing the decline of entanglement. This section focuses on the distinctions between the x--axis and z--axis DM interaction parameters on the LQU.

\begin{figure}[!h]
  \centering
  \includegraphics[width=3.3in]{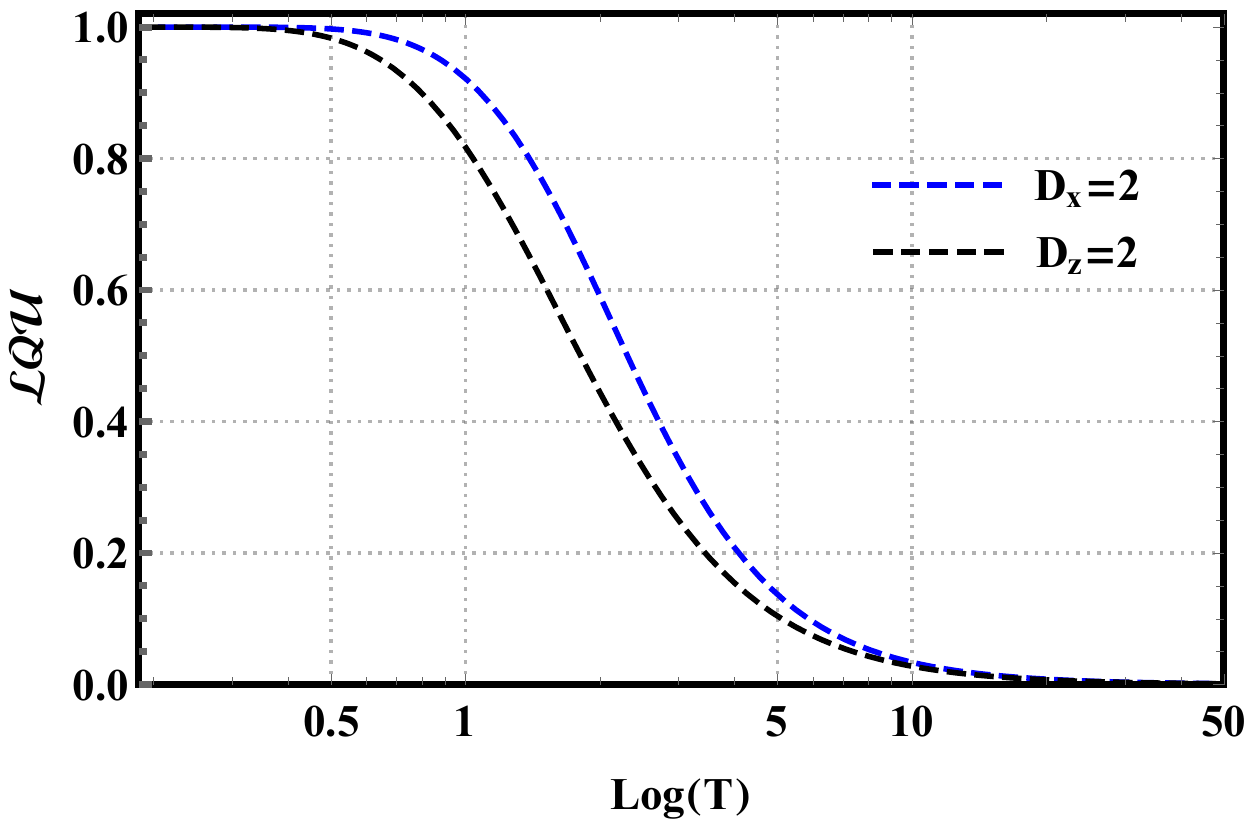}
  \caption{: The LQU is plotted as a function of the logarithmic temperature
$Log(T)$ for $D_z = 2$ and $D_x = 2$, here $J=1$  and $\Delta=0.5$.}\label{f7}
\end{figure}
We can see from graph (\ref{f7}) that raising the x--axis parameter $D_x$ causes the correlation measured by LQU to grow more rapidly. Furthermore, when $Log(T) = 2$, the value of LQU for the $D_x$ parameter (blue dashed curve) is superior to that of $D_z$ (black dashed curve). These phenomena indicate that the x--axis parameter $D_x$ has a more significant effect than the z-axis parameter $D_z$ on the quantum correlations.

\section{Conclusion}
The correlation of a two--qubit Heisenberg XY system with distinct DM interaction parameters has been examined using local quantum uncertainty measurement. The DM interaction and coupling parameter $J$ are excellent correlation control basic features. We may improve the correlation by raising the parameters. Furthermore, we studied the impact of the x--axis and z--axis DM interaction components on the LQU. Correlation may be increased faster by raising $D_x$ rather than $D_z$. When the values of $D_x$ and $D_z$ are the same, as a result, by modifying the direction of the DM interaction, we may achieve a more efficient command parameter for increasing correlation.



\begin{thebibliography}{99}
\bibitem{Hor} R. Horodecki, P. Horodecki, M. Horodecki and K. Horodecki, Rev. Mod. Phys. \textbf{81} 865 (2009).
\bibitem{Niel} M. A. Nielsen and I. L. Chuang, Quantum Computation and Quantum Information (Cambridge University Press, Cambridge, 2000).
\bibitem{Fer} A. Ferraro, L. Aolita, D. Cavalcanti, F. M. Cucchietti and A. Acin, Phys. Rev. A \textbf{81} 052318 (2010).
\bibitem{Mod} K. Modi, A. Brodutch, H. Cable, T. Paterek and V. Vedral, Rev. Mod. Phys. \textbf{84} 1655 (2012).
\bibitem{Hend} L. Henderson and V. Vedral, J. Phys. A : Math. Gen. \textbf{34}  6899 (2001).
\bibitem{Olli} H. Ollivier and W. H. Zurek, Phys. Rev. Lett. \textbf{88}  017901 (2001).
\bibitem{Dak} B. Dakic, V. Vedral and C. Brukner, Phys. Rev. Lett. \textbf{105} 190502 (2010) .
\bibitem{Oste} A. Osterloh, L. Amico, G. Falci and R. Fazio, Nature (London) \textbf{416} 608 (2002).
\bibitem{Amesen} M. C. Amesen, S. Bose, and  V. Vedral, Phys. Rev. Lett. \textbf{87} (2001) 017901
\bibitem{WangX.G} X. G. Wang, Phys. Rev. A 64 (2001) 012313
\bibitem{ZhouL} L. Zhou, H. S. Song, Y. Q. Guo, and C. Li, Phys. Rev. A. \textbf{68} (2003) 024301
\bibitem{ZhangG.F} G. F. Zhang, and S. S. Li, Phys. Rev. A. \textbf{72} (2005) 034302
\bibitem{Guo1} J. L. Guo, and H. S. Song, Eur. Phys. J. D. \textbf{56} (2010) 265
\bibitem{Houca1} R. Houça, A. Belouad, E. B. Choubabi, A. Kamal, and M. El Bouziani, J. Magn. Magn. Mater. \textbf{563} (2022) 169816
\bibitem{Tufarelli} T. Tufarelli, D. Girolami, R. Vasile, S. Bose, and G. Adesso, Phys. Rev. A. S\textbf{86} (2012) 052326
\bibitem{YaoY1} Y. Yao, and al, Phys. Rev. A. \textbf{86} (2012) 062310
\bibitem{Song23} S. Xu, X. K. Song, and L. Ye, Int. J. Mod. Phys. B.  \textbf{27} (2013) 1350074
\bibitem{Yao2} Y. Yao, and al. Phys. Rev. A. \textbf{86} (2012) 042102
\bibitem{Liu23}  F. W. Ma, S. X. Liu, and X. M. Kong, Phys. Rev. A. \textbf{84} (2011) 042302
\bibitem{Wang X G33} X. G. Wang, Phys. Rev. A. \textbf{64} (2001) 012313
\bibitem{Wang X G34} X. G. Wang, Phys. Rev. A. \textbf{66} (2002) 034302
\bibitem{Houca2} R. Houça, A. Belouad, E. B. Choubabi, A. Kamal, and M. El Bouziani, Quant. Infor. Proc. \textbf{21} (2022) 1
\bibitem{Nielsen} M. A. Nielsen,''Quantum information theory'' PhD thesis, University of New Mexico (1998). arXiv:quant-ph/0011036
\bibitem{Arnesen} M. C. Arnesen, S.Bose, and V. Vedral, Phys. Rev. Lett. \textbf{87} (2001) 017901
\bibitem{Kamta G L} G. L. Kamta, and A F. Starace, Phys. Rev. Lett.  \textbf{88} (2002) 107901
\bibitem{Wang X G35} X. G. Wang, Phys. Lett. A. \textbf{281} (2001) 101
\bibitem{Zhou L36} L. Zhou, H. S. Song, Y. Q. Guo, et al. Phys. Rev. A. \textbf{68} (2003) 024301
\bibitem{William K} K. W. William, Phys. Rev. Lett. \textbf{80} (1998) 2245
\bibitem{Wang45} X. Wang, H. B. Li,Z. Sun, et al, J. Phys. A. \textbf{38} (2005) 870
\bibitem{Gir} D. Girolami, T. Tufarelli, G. Adesso, Phys. Rev. Lett. \textbf{110}, 240402 (2013).
\bibitem{Luo} S. Luo, Phys. Rev. Lett. \textbf{91}, 180403 (2003).
\bibitem{Wig} E.P. Wigner, M.M. Yanase, Proc. Natl. Acad. Sci. U.S.A. \textbf{49}, 910 (1963).
\bibitem{Luoo} S. Luo, S. Fu, C.H. Oh, Phys. Rev. A \textbf{85}, 032117 (2012).
\bibitem{mez} M. Ali, Eur. Phys. J. D  \textbf{74} 186 (2020).
\bibitem{Sen} A. Sen, A. Bhar, D. Sarkar, Quan. Info. Proc. \textbf{14}, 269 (2015).
\bibitem{Kar} G. Karpat, B. Cakmak, F.F. Fanchini, Phys. Rev. B \textbf{90}, 104431 (2014).
\bibitem{Coul} I.B. Coulamy, J.H. Warnes, M.S. Sarandy, A. Saguia, Phys. Lett. A \textbf{380}, 1724 (2016).
\bibitem{Wu} S-X. Wu, Y. Zhang, C-S. Yu, Ann. Phys. \textbf{390}, 71 (2018).
\bibitem{Sla1} A. Slaoui, M.I. Shaukat, M. Daoud, R. Ahl Laamara, Eur. Phys. J. Plus \textbf{133}, 413 (2018).
\bibitem{Sla2} A. Slaoui, M. Daoud, R. Ahl Laamara, Quan. Info. Proc. \textbf{17}, 178 (2018).
\bibitem{Sla3} A. Slaoui, L. Bakmou, M. Daoud, R. Ahl Laamara, Phys. Lett. A \textbf{383}, 2241 (2019).
\bibitem{Sla4} A. Slaoui, M. Daoud, R. Ahl Laamara, Quan. Info. Proc. \textbf{18}, 250 (2019).
\bibitem{Jour} New Journal of Physics Focus Issue on Gravitational Quantum Physics (2014)
http://iopscience. iop.org/1367–2630/focus/Focus
\bibitem{Karo} F. Karolyhazy, Nuov. Cim. A\textbf{42}, 390 (1966).
\bibitem{pol} C. Anastopoulos and B. L. Hu, Class. Quantum Grav. \textbf{37}, 235012 (2020).
\bibitem{rojas} M. Rojas, and I. P. Lobo, ''Thermal quantum correlations in two gravitational cat states'', [arXiv:21.06.05696] (2021).
\bibitem{Girolami} D. Girolami, T. Tufarelli and G. Adesso, Phys. Rev. Lett. \textbf{110}, 240402 (2013).
\bibitem{Wang} S. Wang, H. Li, X. Lu and G-L, Long  arXiv:1307.0576v2.
\bibitem{Kedif} Y. Khedif and M. Daoud, Int. J. Mod. Phys. B \textbf{32}, 1850218 (2018).
\bibitem{Kedif1} Y. Khedif and M. Daoud, Quantum Inf. Process. \textbf{18}, 45 (2019).
\bibitem{Jebli} L. Jebli, B. Benzimoun  and M. Daoud,  Int. J. Quantum Inf. \textbf{15}, 1750020 (2017).
\end{thebibliography}
\end{document}